# A rutile-based homologous series Na(PtO$_2$)$_{2n+1}$ discovered by computationally assisted high-pressure synthesis

Yasuhito Kobayashi[1*], Hidefumi Takahashi[1,2], Shunsuke Kitou[2], Akitoshi Nakano[3], Hajime Sagayama[4], Yuichi Yamasaki[5], and Shintaro Ishiwata[1,2*]

[1] *Division of Materials Physics, Graduate School of Engineering Science, Osaka University, Toyonaka, Osaka 560-8531, Japan*
[2] *Spintronics Research Network Division, Institute for Open and Transdisciplinary Research Initiatives, Osaka University, Yamadaoka 2-1, Suita, Osaka 565-0871, Japan*
[2] *Department of Advanced Materials Science, The University of Tokyo, Kashiwa 277-8561, Japan*
[3] *Department of Applied Physics, Nagoya University, Furo-cho, Chikusa-ku, Nagoya 464-8603, Japan*
[4] *Institute of Materials Structure Science (IMSS), High Energy Accelerator Research Organization (KEK), Tsukuba, Ibaraki 305-0801, Japan*
[5] *Center for Basic Research on Materials, National Institute for Materials Science (NIMS), Tsukuba, Ibaraki 305-0047, Japan*

*To whom correspondence should be addressed;
Email: kobayashi@qm.mp.es.osaka-u.ac.jp, ishiwata.shintaro.es@osaka-u.ac.jp

**Abstract**

Layered transition metal oxides typified by the Ruddlesden-Popper phase have been extensively studied for its applications in high-temperature superconductivity, catalysis, and battery technologies. Despite the remarkable structural diversity and catalytic functionality of platinum oxides, the exploration of layered polymorphs has remained significantly constrained mainly due to the high inertness of platinum. Here, we discover a new homologous series of layered ternary oxides, Na(PtO$_2$)$_{2n+1}$, by a combination of highly oxidizing high-pressure methods and density functional theory (DFT) calculations. This series features unprecedented layered structural motifs, rutile-based PtO$_6$ octahedra and one-dimensional PtO$_4$ square-planar columns, which enables systematic control of dimensionality. Furthermore, we demonstrate a computationally-assisted identification of isomeric and putative members of this homologous series as confirmed by controlled synthesis and quantitative analysis of diffuse scattering data. This approach provides an effective platform for the exhaustive exploration of metastable transition metal oxides with rich structural variations.

## I. INTRODUCTION

Layered homologous series of transition metal oxides (TMOs), especially the Ruddlesden-Popper (RP) series $A_{n+1}B_nO_{3n+1}$, have long been at the forefront of materials science due to their remarkable structural and functional versatility. The discovery of high-temperature superconductivity in layered cuprates, followed by ruthenates and nickelates, underscores the pivotal role of RP phases in advancing our understanding of condensed matter physics[1–3]. Beyond superconductivity, RP oxides have demonstrated exceptional catalytic performance in electrocatalytic energy conversion reactions[4–7]. Their high ionic conductivity and remarkable structural stability render them promising candidates for solid-state batteries[8–11]. In such a layered structural family, their alternating perovskite-type and rock-salt-type layers constitute a modular architecture governed by homologous series, wherein variations in the dimensionality enable systematic tuning of their physical and chemical properties[12–14].

While layered homologous series of TMOs have been extensively studied as exemplified by RP series, Dion-Jacobson series, and Aurivillius series[15–18], the development within 5d TMOs has remained limited, largely due to the high inertness of 5d transition metals[19,20]. To date, RP series and related layered series in 5d TMOs have been reported only in a few systems, such as those based on tantalum, tungsten, and iridium[17,18,21–24], highlighting the significant gap in their exploration. Among the 5d TMOs which may exhibit exotic properties arising from the strong spin-orbit coupling, platinum-based systems present an exceptional opportunity for exploration. Platinum compounds are well known for their catalytic excellence, exemplified by their role in oxygen evolution and reduction reactions[25–29], as well as the Adams catalyst ($\alpha$-PtO$_2$), which demonstrates remarkable activity in hydrogenation and dehydrogenation processes critical to energy conversion technologies[30]. Beyond their catalytic capabilities, platinum-based oxides exhibit a unique structural adaptability. Platinum ions can adopt both octahedral coordination (Pt$^{4+}$) and square-planar coordination (Pt$^{2+}$)[31,32], a rare coexistence observed in compounds such as $A$Pt$_3$O$_6$[33–35] and PbPt$_2$O$_4$[36,37]. This dual coordination not only sets platinum oxides apart from the other 5d TMOs but also provides extraordinary structural flexibility, potentially enabling the design of novel layered homologous series.

In this study, we report a new layered homologous series of ternary sodium-platinum oxides, Na(PtO$_2$)$_{2n+1}$ ($n$: number of PtO$_6$ octahedral layers), using a high-pressure synthesis method with highly oxidizing atmosphere. We further demonstrate the predictive power of a density functional theory (DFT)-based computational framework, which enables the efficient exploration and identification of new, stable polymorphs in this system. The crystal structures of Na(PtO$_2$)$_{2n+1}$ series consist of a unique alternate stacking of a rutile-based block of PtO$_6$ octahedra and a single layer of a bunch of one-dimensional square-planar PtO$_4$ units, markedly different from conventional layered homologous series such as the RP phases, which are composed of perovskite-based layered blocks. This unprecedented structural geometry, with controllable dimensionality, would provide a new platform for investigating the structure-property relationships of layered homologous series.

## II. RESULTS AND DISCUSSION

**Synthesis and characterization of NaPt$_3$O$_6$**

As shown in Fig. 1a, needle-like single crystals of NaPt$_3$O$_6$ up to approximately 200 μm in length and less than 1 μm in diameter were obtained upon slowly cooling the sample from 1200 °C to 950 °C under 4 GPa, as detailed in Methods section. Fig. 1b shows the crystal structure of NaPt$_3$O$_6$ at 150 K obtained by synchrotron X-ray diffraction (SXRD) experiments on a single crystal. The SXRD data were indexed by the monoclinic lattice with space group of $P2/m$, as shown in Fig. 1c. The obtained structural parameters are shown in Table 1, yielding reliability factors of $wR_2$ = 4.01 and $S$ = 1.05 using all observed reflections (for detailed refinement statistics, see Table S2). The refined atomic displacement parameters (ADP) for Na ions were anomalously large compared to the lighter O ions. These elevated ADPs for Na ions, which exhibit an anisotropic distribution aligning with their decahedral coordination geometry (Fig. S1, Table S3), are primarily attributed to the increased volume of the decahedra. This high coordination number likely allows for greater vibrational amplitudes, particularly along directions corresponding to longer Na-O bond distances within the distorted polyhedron. Polycrystalline samples of NaPt$_3$O$_6$ were also obtained by high-pressure synthesis as detailed in the Methods section (for detailed refinement, see Fig. S2 and Table S1). It has been reported that some $A$Pt$_3$O$_6$ compounds exhibit nonstoichiometry at the $A$-site[34]; however, our quantitative analysis of the chemical composition for the NaPt$_3$O$_6$ single crystal using energy dispersive X-ray spectroscopy indicates nearly full stoichiometry, with Na and Pt in a molar ratio of 0.96(6) : 3.

The crystal structure of NaPt$_3$O$_6$ includes three non-equivalent Pt sites: two with octahedral coordination (Pt1, Pt2) and one with square-planar coordination (Pt3), as shown in Fig. 1b. PtO$_6$ octahedra form two-dimensional layers, while PtO$_4$ square planes form one-dimensional columnar stacks along the $b$-axis. This crystal structure represents a novel polymorph distinct from the previously reported orthorhombic $A$Pt$_3$O$_6$ counterparts[33-35]. NaPt$_3$O$_6$ features a rutile-based layer extended along the [110] direction of $β$-PtO$_2$[38]. $β$-PtO$_2$ is a typical binary platinum oxide with a rutile-type structure which tends to be stabilized at high pressures. In the framework of an ionic bonding model, the valence states of Pt ions in previously reported $A$Pt$_3$O$_6$ compounds, where $A$ is a divalent cation, can be explained by assigning the oxidation states of +4 to two octahedrally coordinated Pt sites and +2 to one square-planar coordinated Pt site, thereby achieving overall charge neutrality. However, this conventional oxidation-state distribution (+4 for the octahedral sites and +2 for the square-planar site) fails to satisfy charge neutrality in NaPt$_3$O$_6$, which contains monovalent Na ions. In fact, bond valence sum[39] (BVS), an empirical method to estimate ionic valence calculated from metal-oxygen bond distances, calculated from the structural parameters determined via the single-crystalline SXRD analysis, revealed valence states of +4.110(18) for Pt1, +4.135(19) for Pt2 (both octahedral sites), and +2.360(17) for Pt3 (the square-planar site). Note that the BVS of cations in the metastable oxides obtained by high-pressure synthesis tends to be larger than that expected for the nominal oxidation state, reflecting the effect of overbonding[40]. The experimental result that only Pt3 shows a relatively large increase in its oxidation state compared with those in other $A$Pt$_3$O$_6$ may indicate the possibility of a rare trivalent oxidation state for square-planar Pt sites in NaPt$_3$O$_6$. However, given the relatively small electronegativity difference between platinum and oxygen, the bonding likely has significant covalent character, which complicates a straightforward assignment of formal oxidation states based solely on BVS calculations. Therefore, spectroscopic probes like XPS are crucial for verifying oxidation state of platinum. For instance, platinum ion in CaPt$_2$O$_4$ is formally trivalent, yet XPS studies have suggested a divalent oxidation state[34].

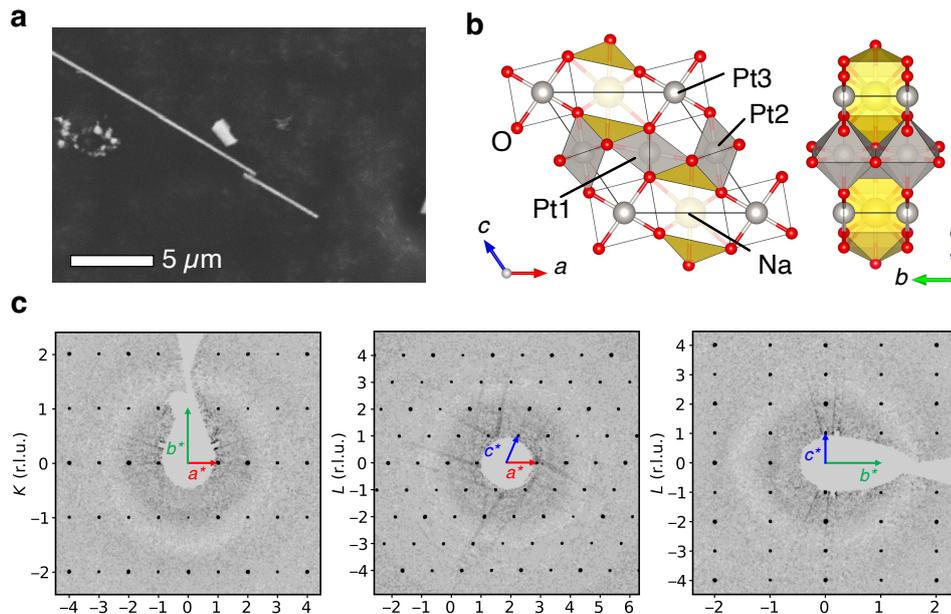

**Fig. 1. Structure characterization of NaPt$_3$O$_6$. a**, Scanning electron microscopy image of the needle-like and platelet shaped single crystals of NaPt$_3$O$_6$. **b**, Crystal structure of NaPt$_3$O$_6$. Three non-equivalent Pt sites are labeled; Pt1 and Pt2 are octahedrally coordinated and Pt3 is planar tetrahedrally coordinated. **c**, Reconstructed precession images of the reciprocal lattice space for $HK$0, $H$0$L$, and 0$KL$ planes, respectively. Diffraction patterns for the single-crystal NaPt$_3$O$_6$ were obtained at 150 K.

**Table 1.** Lattice parameters and atomic coordinates obtained from SXRD analysis of NaPt$_3$O$_6$ single crystal. The SXRD measurement was performed at 150 K using synchrotron radiation with a wavelength of $\lambda$ = 0.3095 Å. Refinement using all 3576 unique reflections yielded reliability factors of $wR_2$ = 4.01 and $S$ = 1.05. Occupancies of all sites are fixed at 1.0.

| Cell | $a$ (Å) | $b$ (Å) | $c$ (Å) | $\beta$ (°) | $V$ (Å$^3$) |
|---|---|---|---|---|---|
| $P2/m$ (No.10) | 6.3724(2) | 3.14060(10) | 7.0327(3) | 124.100(4) | 116.546(9) |

| Atom | $x$ | $y$ | $z$ | $g$ | site |
|---|---|---|---|---|---|
| Na1 | 0.5 | 0.5 | 0 | 1 | 1$f$ |
| Pt1 | 0.5 | 0.5 | 0.5 | 1 | 1$e$ |
| Pt2 | 0 | 0 | 0.5 | 1 | 1$g$ |
| Pt3 | 0 | 0 | 0 | 1 | 1$a$ |
| O1 | 0.7668(4) | 0.5 | 0.4324(4) | 1 | 2$n$ |
| O2 | 0.8364(4) | 0 | 0.1609(3) | 1 | 2$m$ |
| O3 | 0.6416(4) | 0 | 0.7201(3) | 1 | 2$m$ |

**Stability of the monoclinic NaPt$_3$O$_6$ structure compared to orthorhombic $A$Pt$_3$O$_6$ analogues**

Next, we discuss the origin of the stability of the Na(PtO$_2$)$_{2n+1}$ series with rutile-type layers on the basis of an ionic model, where metal cations occupy the interstitial sites between the closely-packed oxygen ions (Fig. 2a,b). Here, we compare the structure of NaPt$_3$O$_6$, corresponding to $n$ = 1, to that of CdPt$_3$O$_6$, which is the representative of the $Cmmm$-type structure of $A$Pt$_3$O$_6$. The previously reported $A$-site cations in $A$Pt$_3$O$_6$ including Cd are divalent unlike Na[34]. As shown in Fig. 2b, the crystal structure of CdPt$_3$O$_6$ can be characterized by the two-dimensional PtO$_6$ octahedra and the strong tendency of Pt ions to adopt square-planar coordination. The two-dimensional layer, where PtO$_6$ octahedra share the edge and the corner along the $b$-axis and the $a$-axis, respectively, is isostructural with that found in CaPtO$_3$ with post-perovskite-type structure[41]. The crystal structure of NaPt$_3$O$_6$, characterized by the rutile-type layer, can be distinguished from the CdPt$_3$O$_6$-type structure from the viewpoint of the connections of PtO$_6$ octahedra. As shown in Fig. 2a, while the PtO$_6$ octahedra form a triangular lattice in the $ab$ plane, those in the CdPt$_3$O$_6$-type structure form a square lattice. The difference in the octahedral connection originates from the fact that the Na ion having a larger ionic radius favors higher coordination numbers as compared with the Cd ion. The position of partial oxygen atoms is modified and the coordination number of Na polyhedra is increased from 8 to 10 (for detailed change in oxygen positions, see Fig. S3a), resulting in the rutile-type octahedral connection for the NaPt$_3$O$_6$ structure.

The stability of the $P2/m$-type structure of NaPt$_3$O$_6$ and $Cmmm$-type CdPt$_3$O$_6$ is also supported by thermodynamic DFT calculations. As shown in Figs. 2c,d, structural relaxation calculations under both ambient and high-pressure conditions were performed for the NaPt$_3$O$_6$ and CdPt$_3$O$_6$ compositions with previously reported orthorhombic structures ($Cmmm$, $Pbam$, and $Pn2_1m$, as shown in Fig. S3b) and the monoclinic structure ($P2/m$). The enthalpies of optimized structures revealed that the most stable structure for CdPt$_3$O$_6$ in the pressure range of 0 to 10 GPa is the previously reported $Cmmm$-type structure. On the other hand, the $P2/m$ structure is thermodynamically stable for NaPt$_3$O$_6$ compared to any other orthorhombic structures, which agrees with the experimental results. To summarize, the stability of rutile-type layered structures, such as that observed in NaPt$_3$O$_6$, depends critically on the ionic radius and coordination preference of the $A$ site cations. Among the $A$-site cations reported for $A$Pt$_3$O$_6$ (inset of Fig. 2c), the Na ion has the largest ionic radius in an 8-fold coordination environment, which explains the stability of the NaPt$_3$O$_6$ structure.

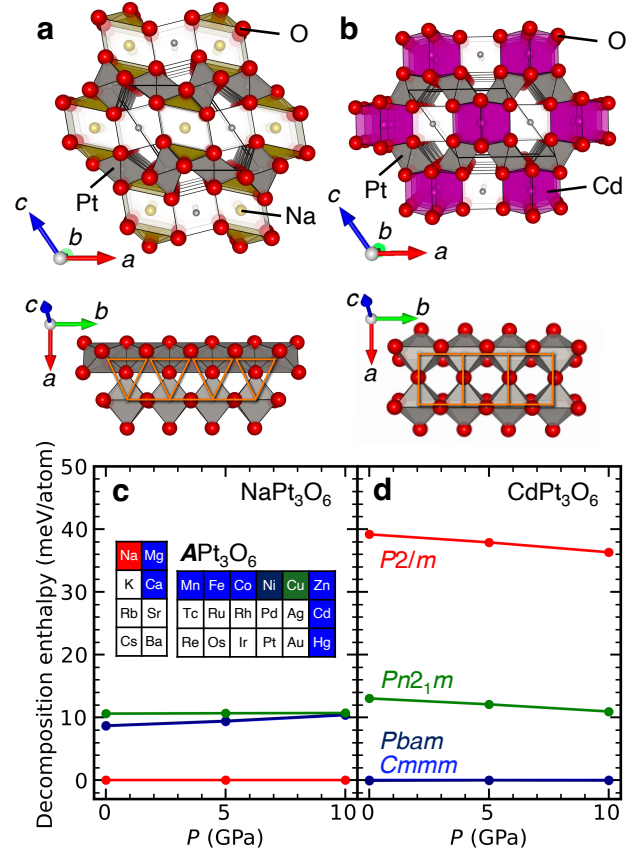

**Fig. 2. Evolution and stability of the polymorph with rutile-based PtO$_6$ layers and columnar stackings of PtO$_4$.** Crystal structures of (a) NaPt$_3$O$_6$ and (b) CdPt$_3$O$_6$, where each atom is shown as an ionic rigid sphere. For clarity, the primitive unit cell of CdPt$_3$O$_6$ is displayed instead of the conventional *Cmmm* unit cell (see Fig. S3). The geometrical differences between the two structures are highlighted by the connection of PtO$_6$ octahedra, where NaPt$_3$O$_6$ shows the rutile-type connection and CdPt$_3$O$_6$ shows the post-perovskite-type connection. Orange lines indicate triangular and square lattices in the *ab* plane formed by PtO$_6$ octahedra for NaPt$_3$O$_6$ and CdPt$_3$O$_6$, respectively. Pressure dependence of decomposition enthalpies for (c) NaPt$_3$O$_6$ and (d) CdPt$_3$O$_6$ with 4 relaxed structures. Inset shows the cation reported in *A*Pt$_3$O$_6$, each corresponding to the color of the plot.

**Computational predictions for a novel layered homologous series Na(PtO$_2$)$_{2n+1}$**

Layered oxides composed of transition metal octahedra frequently constitute a homologous series with variable layer number as stable phases, such as RP phases. To search for the platinum-based layered oxides with various numbers of the rutile-type layers, we performed DFT calculations for the thermodynamic convex hull of the Na-Pt-O ternary system under high pressures, as shown in Fig. 3. We define a hypothetical layered homologous series with the composition Na(PtO$_2$)$_{2n+1}$, where $n$ corresponds to the number of rutile-type layers, with $n = 1$ corresponding to the experimentally-obtained phase NaPt$_3$O$_6$. It should be noted that this hypothetical series can be extended to the infinite-layer phase ($n = \infty$), that is $\beta$-PtO$_2$. The formation enthalpy for NaPt$_3$O$_6$ ($n = 1$) was calculated by performing structural relaxations using the experimentally refined crystal structure as an initial model. As shown in Fig. 3b, we manually generated and optimized for four hypothetical members of Na(PtO$_2$)$_{2n+1}$: NaPt$_5$O$_{10}$ ($n = 2$) as a double-layer phase, NaPt$_7$O$_{14}$ ($n = 3$) as a triple-layer phase, NaPt$_9$O$_{18}$ ($n = 4$) as a quadruple-layer phase, and NaPt$_4$O$_8$ ($n = 1.5$) as an alternate-stacking phase of single-layer and double-layer ones. In the structural relaxation calculations, each of the structural parameters converged to a reasonable configuration under space group $P2/m$ similar to that of NaPt$_3$O$_6$.

The thermodynamic convex hull calculated at ambient pressure (0 GPa) confirmed the reproducibility of DFT-based calculations, as all known compounds in the Na-Pt-O ternary system occupied the vertices of the convex hull (Fig. S4). The convex hull calculated at 5 GPa (Fig. 3a), close to the experimental high-pressure conditions in this study, demonstrates that NaPt$_3$O$_6$ occupies a vertex of the convex hull, confirming its stability at high pressure and aligning with experimental results. Hypothetical members with $n$ = 2, 3, 4, and 1.5 are above the convex hull, indicating they are metastable. This metastability at high pressure can be associated with the stability of the closely related compound $\beta$-PtO$_2$ characterized by dense structure[42].

To visualize the compositional range of Na(PtO$_2$)$_{2n+1}$ and pressure dependence of the convex hull, we show the cross-section of the ternary convex hull, focusing on the compositional range of Na-PtO$_2$ (Fig. 4a), and the pressure dependence of the decomposition enthalpy for Na(PtO$_2$)$_{2n+1}$ (Fig. 4b). Our calculations indicate that the decomposition enthalpies of metastable phases with $n$ = 2, 3, 4, and 1.5 are all less than 5 meV/atom. It should be noted that DFT-based energy calculations can generally deviate by several tens of meV/atom due to the omission of entropy and approximate functionals[43–46]. Consequently, the hypothetical members can be regarded as nearly stable phases, indicating experimental accessibility of the layered homologous series Na(PtO$_2$)$_{2n+1}$. The increase of the decomposition enthalpies of further Na(PtO$_2$)$_{2n+1}$ members at elevated pressures, indicating that high-pressure techniques are not suitable for obtaining these members. However, high-pressure techniques with oxidizers are necessary because platinum-based oxides are generally prone to reduction to elemental Pt at ambient pressure, especially above 1000 °C[20]. Considering the extremely high oxygen pressure conditions and the thermodynamic stability at high pressures, the optimal pressure for further stabilizing the other members of the Na(PtO$_2$)$_{2n+1}$ series should be located at moderate pressures as 4 GPa, which was adopted for the single-layer phase NaPt$_3$O$_6$.

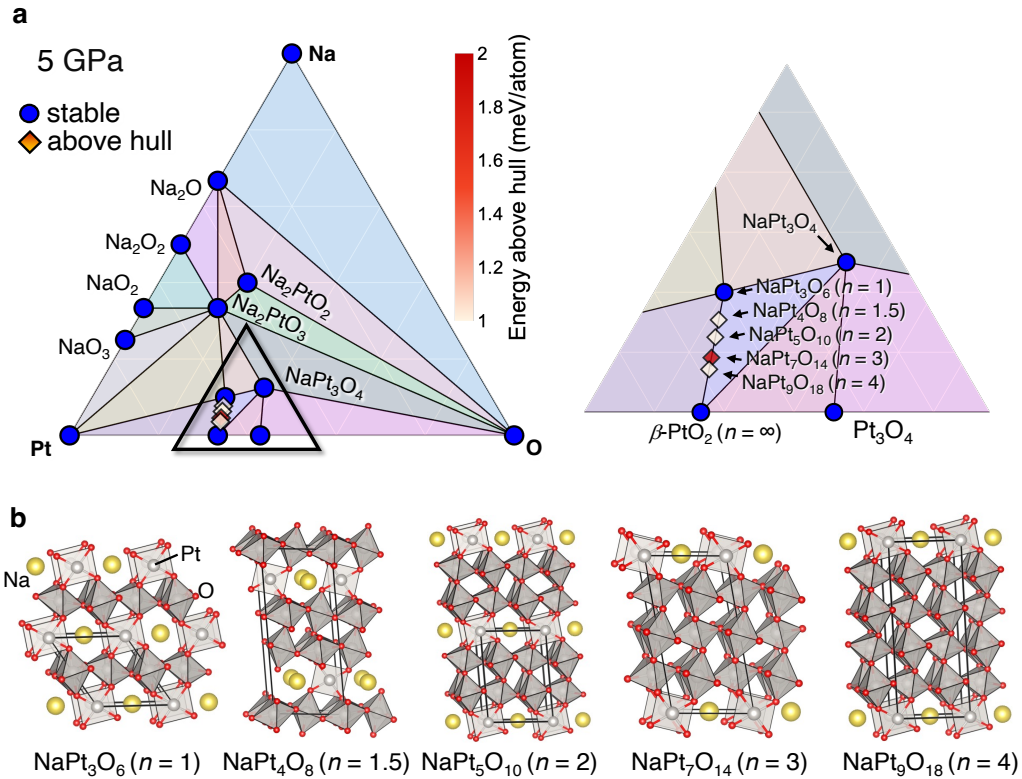

**Fig. 3. DFT-based Predictions for members of Na(PtO$_2$)$_{2n+1}$ homologous series. a**, Convex hull diagram for Na-Pt-O ternary system under 5 GP (left). Black solid lines show the convex hull, blue circles show stable phases, and square markers show metastable phases with a color scale of energy above the convex hull. Closed view of the convex hull indicated by black triangle (right). **b**, Crystal structures of Na(PtO$_2$)$_{2n+1}$ homologous series. Structures except $n$ = 1 are relaxed using DFT calculations from manually-generated configurations.

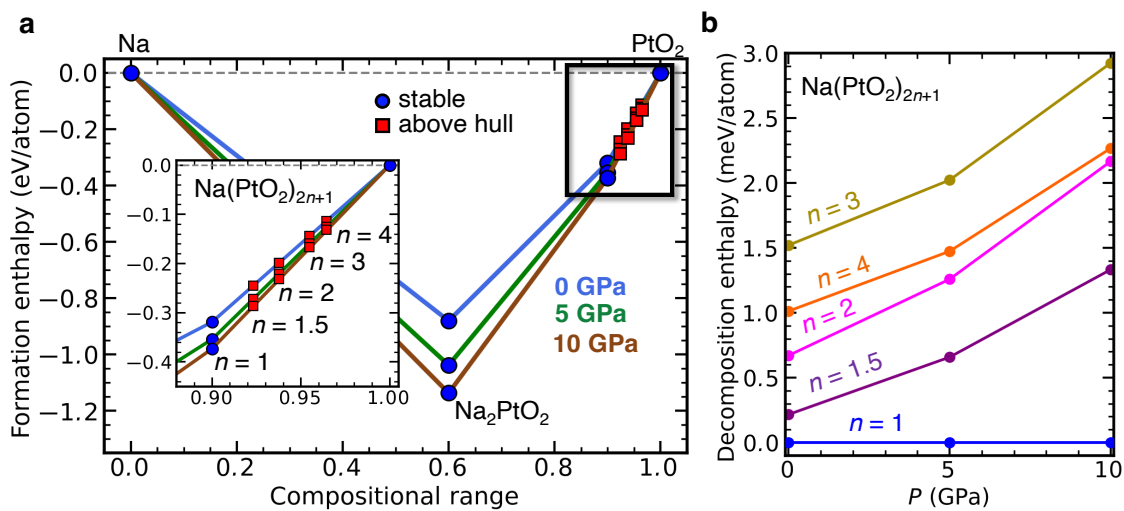

**Fig. 4. Pressure dependence of the stability of Na(PtO$_2$)$_{2n+1}$ homologous series. a**, Pseudo-binary convex hull diagram at various pressures from Na to PtO$_2$. Inset shows closed view of the convex hulls including Na(PtO$_2$)$_{2n+1}$ indicated by the black rectangle. **b**, Pressure dependence of decomposition enthalpies of Na(PtO$_2$)$_{2n+1}$ from 0 GPa to 10 GPa.

**Prediction-based synthesis and characterization of NaPt$_5$O$_{10}$**

Based on the predicted stability of the hypothetical members of Na(PtO$_2$)$_{2n+1}$, we conducted high-pressure synthesis for the double-layer phase NaPt$_5$O$_{10}$ at the moderate high pressure of 4 GPa. Given the nominal oxidation state of Pt ions in NaPt$_5$O$_{10}$ (+3.8) expected to be higher than that in NaPt$_3$O$_6$ (+3.67), it was deemed advisable to slightly decrease the synthesis temperature to mitigate the reducing atmosphere. Sintering conditions at 1150 °C, incorporating the oxidizer KClO$_4$, yielded approximately 10 × 20 × up to 500 μm³ single crystals of NaPt$_5$O$_{10}$ from a product that mainly contained polycrystals of $β$-PtO$_2$ and NaPt$_3$O$_6$, as shown in Figs. 5a,b. The single crystal was characterized by SXRD, as shown in Fig. 5d and refined structural parameters are shown in Table 2. The crystal structure of NaPt$_5$O$_{10}$ shown in Fig. 5c was refined by assuming the monoclinic space group $P2/m$, which is consistent with the predicted phase by DFT, yielding error factors of $wR_2$ = 2.96 and $S$ = 1.56 (for detailed refinement statistics, see Table S4). As shown in Fig. S1 and Table S5, ADPs obtained from single-crystal SXRD revealed an anomalously large value for the Na ions compared to the lighter O ions. This behavior is likely attributable to similar factors discussed for NaPt$_3$O$_6$, namely, the enlarged coordination polyhedron volume associated with the high coordination number. The ADPs of Na ions in NaPt$_5$O$_{10}$ aligns with the local atomic geometry of the decahedral coordination, as found in NaPt$_3$O$_6$. Unlike $A$Pt$_3$O$_6$ having analogues compounds, NaPt$_5$O$_{10}$ represents a new polymorph, featuring the three non-equivalent Pt sites, with octahedral coordination at the Pt1 and Pt2 sites and planar tetrahedral coordination at the Pt3 site. Energy dispersive X-ray spectroscopy on the NaPt$_5$O$_{10}$ single crystal indicates almost no deficiency at Na site, with Na and Pt ions in a molar ratio of 0.98(1) : 5.00(1). BVS calculations based on the structural parameters determined via SXRD analysis reveal the valence states of +4.128(5) for Pt1, +4.162(5) for Pt2 (both octahedral sites), and +2.310(6) for Pt3 (the square-planar site). The significant upward deviation of the oxidation state of Pt3 from +2 may indicate a partial increase in its oxidation state as discussed for the square-planar coordinated Pt site in NaPt$_3$O$_6$. The general formula of Na(PtO$_2$)$_{2n+1}$ can be expressed as NaPt(Pt)$_{2n}$O$_{4n+2}$, by separating the octahedral and square planar Pt sites. Considering the charge neutrality and the strong tendency that the octahedral Pt ions favor +4 state as supported by the BVS calculations, the oxidation state of the square planar Pt site could potentially be +3. However, as discussed for NaPt$_3$O$_6$, it's also possible that the covalent bonding character is not negligible and the actual oxidation state could in fact be +2.

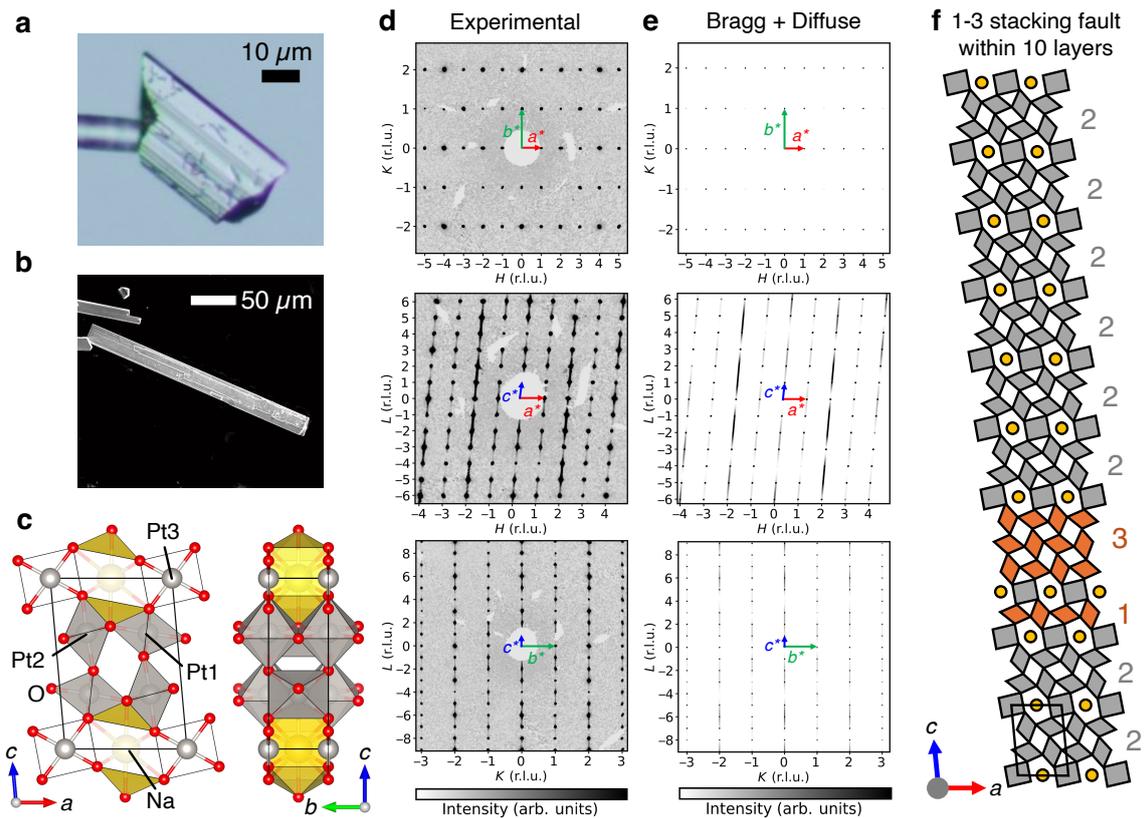

**Fig. 5. Structure and X-ray diffraction patterns of NaPt$_5$O$_{10}$.** **a**, Photograph of the single crystal of NaPt$_5$O$_{10}$ used for SXRD experiments. **b**, Scanning electron microscopy image of NaPt$_5$O$_{10}$ single crystal. **c**, Crystal structure of NaPt$_5$O$_{10}$ at 300 K. Three non-equivalent Pt sites are labeled; Pt1 and Pt2 are octahedrally coordinated and Pt3 is planar tetrahedrally coordinated. **d**, Reconstructed precession images of the reciprocal lattice space for $HK$0, $H$0$L$, and 0$KL$ planes, respectively. Diffraction patterns for the single-crystal NaPt$_5$O$_{10}$ were obtained at 300 K. **e**, Diffraction patterns simulated with a structure introducing stacking faults in the supercell. Structure factors were calculated for $HK$0, $H$0$L$, and 0$KL$ planes, respectively. **f**, Schematic crystal structure used for simulation including stacking faults. Ten $n$ = 2 rutile-type layers were considered for the simulation, two of which are $n$ = 1 and $n$ = 3 continuous defects (1-3-type stacking fault).

Table 2 | Lattice parameters and atomic coordinates obtained from SXRD analysis of NaPt$_5$O$_{10}$ single crystal. The SXRD measurement was performed at 300 K using synchrotron radiation with a wavelength of λ = 0.3094 Å. Refinement using all reflections yielded reliability factors of $wR_2$ = 2.96 and $S$ = 1.56. Occupancies of all sites are fixed at 1.0.

| Cell | a (Å) | b (Å) | c (Å) | β (°) | V (Å³) |
|---|---|---|---|---|---|
| *P*2/*m* (No.10) | 6.40796(5) | 3.14420(2) | 9.07405(7) | 94.9682(7) | 182.136(2) |
| **Atom** | **x** | **y** | **z** | **g** | **site** |
| Na1 | 0.5 | 0.5 | 0 | 1 | 1*c* |
| Pt1 | 0.763393(7) | 0.5 | 0.677790(5) | 1 | 2*m* |
| Pt2 | 0.260571(7) | 0 | 0.680341(5) | 1 | 2*n* |
| Pt3 | 0 | 0 | 0 | 1 | 1*h* |
| O1 | 0.45866(16) | 0.5 | 0.71847(14) | 1 | 2*m* |
| O2 | 0.79003(19) | 0 | 0.82083(13) | 1 | 2*n* |
| O3 | 0.06710(17) | 0.5 | 0.64637(15) | 1 | 2*m* |
| O4 | 0.2888(2) | 0 | 0.45878(13) | 1 | 2*n* |
| O5 | 0.25121(19) | 0 | 0.89722(14) | 1 | 2*n* |

**Origin of the diffuse scattering for NaPt$_5$O$_{10}$**

Unlike NaPt$_3$O$_6$, Bragg reflections indexed by *H*0*L* and 0*kl* in NaPt$_5$O$_{10}$ accompany non-uniform diffuse scattering along the *l* direction. The intensity of the diffuse scattering is approximately three orders of magnitude lower than that of the primary Bragg reflections (see Fig. S5), suggesting it is considered unlikely to have a significant effect on the analysis of the average structure of NaPt$_5$O$_{10}$. The diffuse scattering is observed as a one-dimensional streak extended along the *c*\*-axis in the reciprocal lattice space (see Fig. 5d), indicating the presence of the disorder along the *c* axis. The *l*-dependent distribution of the diffuse scattering intensity suggests that the planar defects are not randomly distributed but have short-range correlations along the *c*-axis. The diffuse scattering likely originates from the Pt ions, the heavy element in the structure, with the planar defects presumably associated with the octahedrally coordinated Pt1 and Pt2 that constitute the two-dimensional layers. The diffuse scattering intensity is relatively stronger when *H* is even in the *H*0*L* plane. The simulated structural factors for NaPt$_5$O$_{10}$ reveal that the contributions from Pt1 and Pt2 constructively interfere when *h* is even and cancel out when *h* is odd (see Fig. S6). In addition, the contribution from Pt3 has almost no *h*-dependency. These results support the origin of diffuse scattering from the two-dimensional Pt-O layers. One possible defect pattern involves the short-range order of stacking faults along the *c*-axis, where single or triple octahedral layers intergrow, which is commonly observed in RP oxides[47–50].

To understand the origin of the diffuse scattering observed in NaPt$_5$O$_{10}$, we performed simulations of the diffuse scattering patterns by introducing partial stacking faults into the supercell and calculating the structural factors. Fig. 5e shows simulated Bragg reflections and diffuse scattering patterns from a crystal structure with 10-layer supercell in which single-layer and triple-layer defects are introduced next to each other (here we call this 1-3-type stacking fault; see Fig. 5f). The simulated diffuse scattering patterns in Fig. 5e are qualitatively consistent with the experimentally observed Miller-index dependence, suggesting possible short-range correlation of the 1-3-type stacking fault along the *c*-axis. Further simulations with different supercell sizes indicate that the Miller-index dependence of the diffuse scattering is invariant and only the relative intensity changes. As a result of the further simulations, the relative intensity to the Bragg reflections was found to be qualitatively in agreement with experiment when one pair of 1-3-type stacking faults is present per approximately 50 layers (see Fig. S7). Other possible stacking fault patterns, such as 1-1-type and 3-3-type layers, reproduced the experimental diffuse scattering less accurately compared to the 1-3-type stacking fault, as shown in Fig. S8. Therefore, it is reasonable to conclude that the diffuse scattering arises from stacking faults of the PtO$_6$ layers, with the 1-3-type stacking fault being the main factor. Fig. S9 shows the decomposition enthalpies calculated by DFT for the 2-2-type and 1-3-type structures of NaPt$_5$O$_{10}$ (without stacking fault), revealing that the 2-2-type structure is stable but the difference between them is less than 1 meV/atom. Therefore, it can be reasonably assumed that 1-3-type stacking faults form within the 2-2-type structure phase. To gain more definitive insights into the local structure of stacking faults, real-space imaging techniques such as transmission electron microscopy would be highly valuable.

## CONCLUSION

In this study, we have successfully synthesized the novel layered oxides NaPt$_3$O$_6$ and NaPt$_5$O$_{10}$ by effectively combining high-pressure syntheses and DFT calculations. NaPt$_3$O$_6$ exhibits a distinct polymorphic structure as a new member of $A$Pt$_3$O$_6$. Building upon the insights gained from the NaPt$_3$O$_6$ structure, we employed DFT calculations to predict the stability of higher-order layered structures and successfully synthesized NaPt$_5$O$_{10}$. This computationally-assisted synthesis unveiled a novel type of layered homologous series Na(PtO$_2$)$_{2n+1}$ with $n = \infty$ representing the well-known rutile-type oxide $\beta$-PtO$_2$. This series showcases quasi-two-dimensional rutile-type PtO$_6$ octahedral layers with variable number of layers and quasi-one-dimensional columnar stacking composed of PtO$_4$ square planes, which potentially cause stacking faults arising from the intergrowth of single and triple layers next to each other. These newly discovered layered structures also open up intriguing possibilities for future investigations into their magnetic properties. Moreover, our computationally assisted approach provides a framework for the comprehensive exploration of homologous series in layered oxides. In contrast to conventional DFT-based convex hull analyses that require prior structural knowledge, this approach overcomes the limitation of this convex hull analysis by initiating from a known layered structure and systematically predicting the stability of unexplored structural homologues.

## METHODS

**Sample preparation**

Single crystals of NaPt$_3$O$_6$ and NaPt$_5$O$_{10}$, and polycrystalline samples of NaPt$_3$O$_6$ were prepared by the solid-state reaction with a high-pressure technique using a cubic-anvil-type high-pressure apparatus. To prepare a precursor for high-pressure synthesis, Na$_2$CO$_3$ (99%, High Purity Chemicals) and PtO$_2$ (99.9%, Sigma-Aldrich) were weighed and mixed in a stoichiometric ratio of 1:1 and heated at 750 °C for 24 hours to release carbon oxides, yielding an orange powder of Na$_2$PtO$_3$[52]. The precursor was then mixed with Pt powder (99.9%, High Purity Chemicals) in a 1.5:1 ratio, and then mixed with the KClO$_4$ powder (20 mg, 99.5%, Wako) as an oxidizer. The excess Na in the mixture compensates for the reaction of Na with the Pt capsule during the synthesis process[53]. The mixture was encapsulated in Pt capsules (bottom diameter 3.8 mm and top diameter of 4.0 mm) and then loaded into a pyrophyllite cube with a dimension of 13 × 13 × 13 mm$^3$ as a pressure medium. The reaction under high pressure was carried out by applying current through a cylindrical graphite heater (5.5 mm in diameter) via Mo electrodes. A cylindrical BN sleeve (6.5 mm in diameter) was inserted between the Pt capsule and the heater. The temperature during the reaction was measured using a thermocouple (Pt wire and Rh-13% Pt wire). Polycrystalline samples of NaPt$_3$O$_6$ were obtained by heating the mixture containing the precursor at 4 GPa and 1200 °C for 30 min., followed by quenching to room temperature before the pressure was released. Powder samples of NaPt$_3$O$_6$ were obtained as a nearly single-phase containing a small amount of NaPt$_3$O$_4$. Single crystals of NaPt$_3$O$_6$ were obtained by heating the same starting materials mixed with a NaCl-KCl flux (twice the weight ratio of the starting material) up to 1200 °C at 4 GPa, followed by cooling to 950 °C in 7 hours. Single crystals of NaPt$_5$O$_{10}$ were obtained by heating the starting materials used for NaPt$_3$O$_6$ at 4 GPa and 1150 °C for 30 min, followed by quenching to room temperature. The product included single crystals of NaPt$_5$O$_{10}$ and polycrystalline PtO$_2$ and NaPt$_3$O$_6$. Note that the products including NaPt$_3$O$_6$ and NaPt$_5$O$_{10}$ are hygroscopic and air-sensitive, likely due to the byproduct of Na$_2$O or Na$_2$O$_2$. The obtained products were then washed and dried with distilled water to remove soluble NaCl, KCl, and the sodium oxides. Single crystals of NaPt$_3$O$_6$ and NaPt$_5$O$_{10}$ are thought to have grown with byproducts of KCl (decomposed from KClO$_4$,) and Na$_2$O or Na$_2$O$_2$ (due to excess amount of Na) acting as a flux.

**Characterization**

Powder XRD experiments using Cu-K$\alpha$ radiation (BRUKER new D8) were performed for phase identification and impurity check. Diffraction data for crystal structure refinement were obtained by powder SXRD experiments at beamline BL-8B of the Photon Factory, KEK in Japan. The X-ray beam was monochromatized using a Si(111) double-crystal monochromator to 18 keV and the wavelength of the beam was determined to 0.6911 Å, which was calibrated with a standard powder sample of CeO$_2$. The monochromatized beam was then focused using a cylindrical bended Rh-coated mirror, and was narrowed down to 0.2 mm using a collimator. The NaPt$_3$O$_6$ powder sample was thoroughly ground using a mortar and pestle, and was subjected to sedimentation in ethanol to achieve uniform particle size. Subsequently, it was loaded into a glass capillary (0.1 mm diameter, 0.01 mm wall thickness, Hilgenberg). The two-dimensional diffraction image was recorded on a cylindrical imaging plate, and then integrated as a function of $2\theta$. The crystal structure was refined by the Rietveld method using the RIETAN-FP software[54]. Rietveld analysis was conducted over the $2\theta$ range from 3° to 180°, excluding the $2\theta$ ranges corresponding to the two main peaks of the impurity NaPt$_3$O$_4$ from the analysis. The powder SXRD pattern indicates the predominant formation of NaPt$_3$O$_6$, with a minor impurity phase of NaPt$_3$O$_4$. Previous investigations on ternary platinum oxides $A$Pt$_3$O$_6$, where counterion $A$ is alkali earth metals, 3d transition metals, and cadmium, have reported orthorhombic crystal structures[35,55,56]. These structures were initially used as starting models for the Rietveld refinement, considering the similarity in diffraction patterns. However, subtle peak splitting could not be adequately resolved, resulting in the enhancement of $R_{wp}$ factor exceeding 7 % when employing the commonly reported *Cmmm* crystal structure. Consequently, the analysis advanced using a symmetry-reduced monoclinic space group, converging on the crystal structure shown in Fig. 1b, with improved error factors of $R_{wp}$ = 1.37 % and $S$ = 1.17.

Single-crystal XRD experiments using Mo-Kα radiation (XtaLAB mini II, Rigaku) were performed for phase identification. Single-crystal SXRD data for crystal structure refinement were obtained at BL02B1 beamline of SPring-8, Japan. The $NaPt_3O_6$ single crystal with the dimensions of approximately 20 μm in length and less than 1 μm in diameter, and the $NaPt_5O_{10}$ single crystal with the dimensions of approximately 30 × 25 × 10 μm$^3$ were employed to obtain diffraction data. The X-ray beam was monochromatized to 40 keV by double-crystal monochromator, and focused using a Pt-coated total-reflection mirror. The wavelength of the beam was 0.3094-0.3095 Å. Diffraction patterns were recorded in the range of $\omega$ = 0-180° using the two-dimensional detector CdTe PILATUS, with 7200 frames for $NaPt_3O_6$ and 720 frames for $NaPt_5O_{10}$. The intensities of Bragg reflections ($d$ > 0.28 Å) were collected by CrysAlisPro program using the fine slice method. Reconstructed precession images of the reciprocal lattice space were obtained with the resolution 0.28 Å. Intensities of equivalent reflections were averaged and the structural parameters were refined by using Jana2006 program[56,58]. Crystal structures were visualized using VESTA software[59]. Simulations of diffuse scattering were performed using the crystal structure factor calculation utility in VESTA. Supercell structures were generated based on structural parameters and anisotropic ADPs obtained from the average structure analysis of $NaPt_5O_{10}$, using Python scripting and VESTA. Crystal structure factors were calculated under λ = 0.3094 Å conditions, incorporating all Na, Pt, and O ions, with anomalous dispersion effects considered.

Scanning electron microscopy and energy dispersive X-ray spectroscopy were performed to investigate morphology and chemical composition of the samples using JCM-7000 NeoScope (JEOL). Samples were placed on carbon tape to decrease charge build-up.

**DFT-based calculations**

The structural relaxations to enthalpically low crystal structures at different pressures were performed within DFT using Quantum Espresso packages[60,61]. Perdew-Burke-Ernzerhof-type (PBE) generalized gradient approximation (GGA) to the exchange-correlation functional and projector-augmented-wave-type (PAW) pseudo-potentials was employed. A plane-wave basis set with a cutoff energies of 100 Ry for wave functions and 800 Ry for charge density were employed to expand the wave functions with the PAW-type pseudo-potentials obtained from the Standard Solid State Pseudopotentials library (SSSP Precision)[62,63]. The reciprocal lattice grid with a spacing of 0.15 Å$^{-1}$ was employed.

For the construction of convex hull diagram, previously reported Na-Pt-O binary and ternary compounds ($Na_2O$, $Na_2O_2$, $NaO_2$, $NaO_3$, $PtO_2$, $Pt_3O_4$, $Na_2PtO_3$, $NaPt_3O_4$, $Na_2PtO_2$, and $Na_4PtO_4$) and terminal compounds (Na, Pt, and O) were subjected to DFT-based structural optimization calculations using existing structural parameters under 0, 5, and 10 GPa. The structural optimization calculations of $NaPt_3O_6$ were performed using refined structural parameters as initial parameters, which were obtained from powder SXRD analysis. The enthalpically low configurations of hypothetical $n$ = 2, 3, 4, 1.5 phases ($NaPt_5O_{10}$, $NaPt_7O_{14}$, $NaPt_9O_{18}$, and $NaPt_4O_8$) were explored by starting from manually generating each crystal structure with the visual support of VESTA. As shown in Fig. S10, each initial configuration was inspired by the layered structure of $NaPt_3O_6$ and generated to satisfy the symmetry of the space group *P*2/*m*, due to the fact that all phases of the layered homologous series, excluding the infinite-layer phase, crystallize in the same symmetry (*I*4/*mmm* in the RP phase with no octahedral distortion, for instance). Note that these initial configurations are not necessarily structurally reasonable, as the bond lengths may be excessively long or short. These structures were enthalpically optimized twice. The first optimizations were performed with low energy cutoffs (66 Ry for wave functions and 432 Ry for charge density) to roughly search for stable configurations. Given the potential for significant changes in lattice dimensions and atomic positions after the initial optimization from manually-generated configurations, the converged structural parameters from the first optimization were employed as second initial parameters and second optimizations were then performed with resampled reciprocal lattice grids and increased cutoff energies. Computations of formation enthalpies for optimized structures and construction of ternary convex hulls were performed using Pymatgen codes implemented in Python[64,65].

**ACKNOWLEDGMENTS**

This work was partly supported by JSPS, KAKENHI (Grants No. 22H00343, No. 23H04871, No. 24K00570, No. 24K17006, No. 24H01644, No. 24KJ1654, and No. 25H00420), FOREST (Grant No. JPMJFR236K) and CREST (Grant No. JPMJCR2435) from JST. The synchrotron radiation experiments were performed at KEK with the approvals of the Photon Factory Program Advisory Committee (Proposal No. 2021S2-004), and at SPring-8 with the approval of the Japan Synchrotron Radiation Research Institute (JASRI) (Proposal No. 2023B1603 and 2024B1842).

**SUPPORTING INFORMATION**

Visualized atomic displacements, Rietveld refinement of powder XRD data of NaPt$_3$O$_6$, previously reported $A$Pt$_3$O$_6$ polymorph, Na-Pt-O ternary convex hull diagram for 0 GPa, structure factor and diffuse scattering analysis of single-crystal XRD data of NaPt$_5$O$_{10}$, enthalpy calculations of Na(PtO$_2$)$_{2n+1}$ series by DFT, and tables showing refined crystallographic parameters and refinement statistics (PDF).

# Supporting information

# A rutile-based homologous series Na(PtO$_2$)$_{2n+1}$ discovered by computationally assisted high-pressure synthesis

Yasuhito Kobayashi[1*], Hidefumi Takahashi[1,2], Shunsuke Kitou[2], Akitoshi Nakano[3], Hajime Sagayama[4], Yuichi Yamasaki[5], and Shintaro Ishiwata[1,2*]

[1] *Division of Materials Physics, Graduate School of Engineering Science, Osaka University, Toyonaka, Osaka 560-8531, Japan*

[2] *Spintronics Research Network Division, Institute for Open and Transdisciplinary Research Initiatives, Osaka University, Yamadaoka 2-1, Suita, Osaka 565-0871, Japan*

[2] *Department of Advanced Materials Science, The University of Tokyo, Kashiwa 277-8561, Japan*

[3] *Department of Applied Physics, Nagoya University, Furo-cho, Chikusa-ku, Nagoya 464-8603, Japan*

[4] *Institute of Materials Structure Science (IMSS), High Energy Accelerator Research Organization (KEK), Tsukuba, Ibaraki 305-0801, Japan*

[5] *Center for Basic Research on Materials, National Institute for Materials Science (NIMS), Tsukuba, Ibaraki 305-0047, Japan*

*To whom correspondence should be addressed;
Email: kobayashi@qm.mp.es.osaka-u.ac.jp, ishiwata.shintaro.es@osaka-u.ac.jp

**Supplementary Figures and Tables**

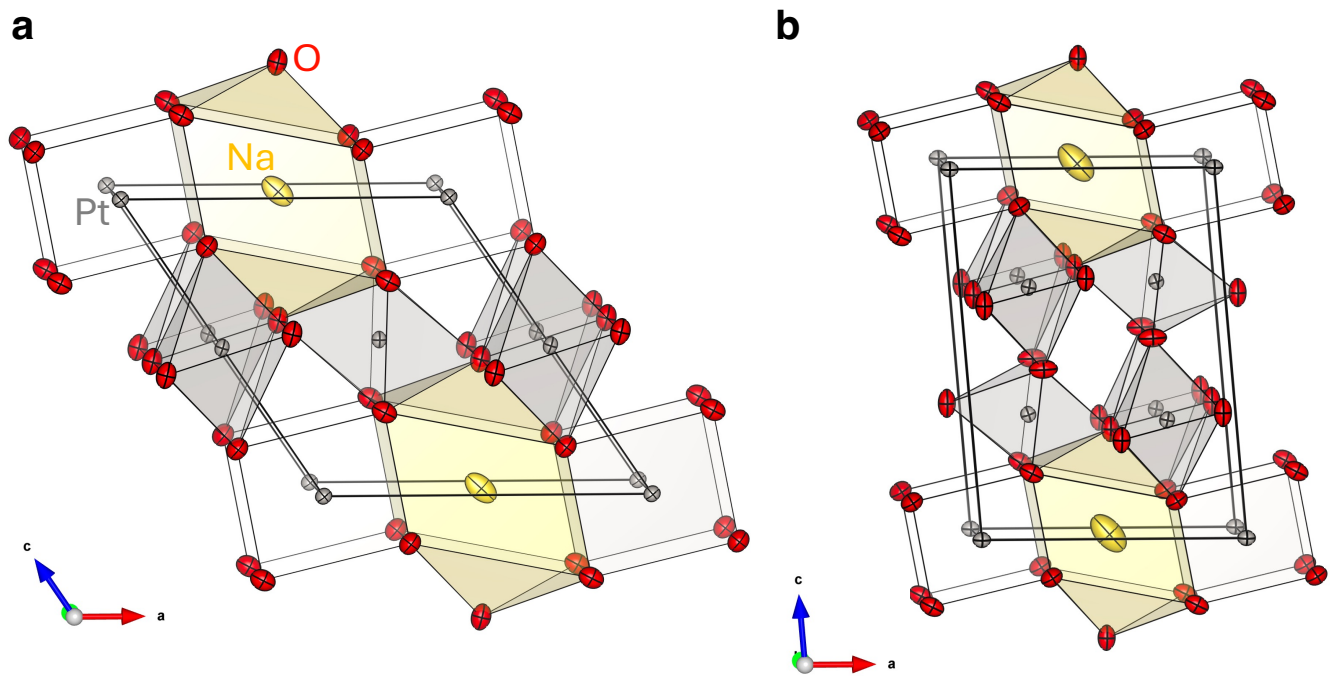

**Fig. S1. Crystal structures of (a) NaPt$_3$O$_6$ at 150 K and (b) NaPt$_5$O$_{10}$ at 300 K obtained by single-crystal SXRD experiments.** Each atom is displayed by displacement ellipsoids (99 % probability).

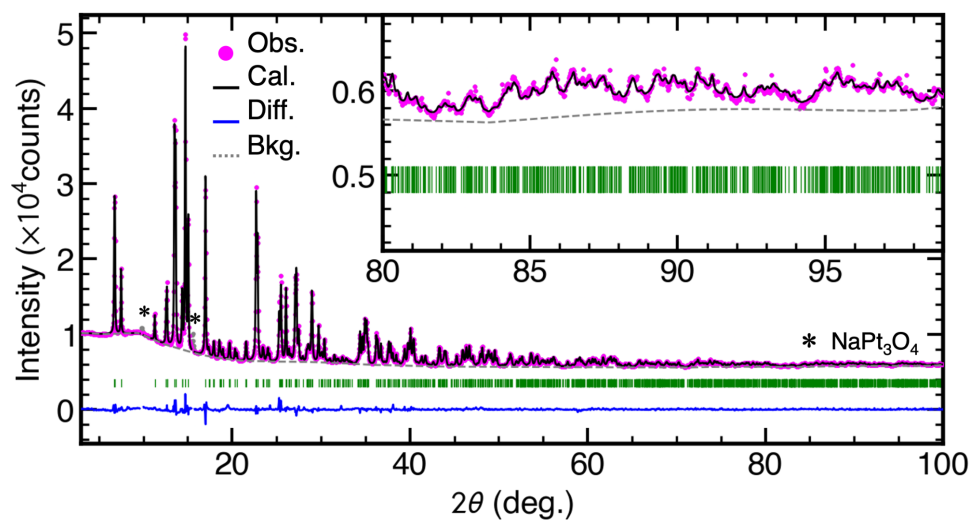

**Fig. S2. Rietveld refinement of powder SXRD data of NaPt$_3$O$_6$, collected using synchrotron radiation with a wavelength of λ = 0.6911 Å.** Bragg reflections are shown as green markers. The refinement yielded error factors of $R_{wp}$ = 1.37 % and $S$ = 1.17.

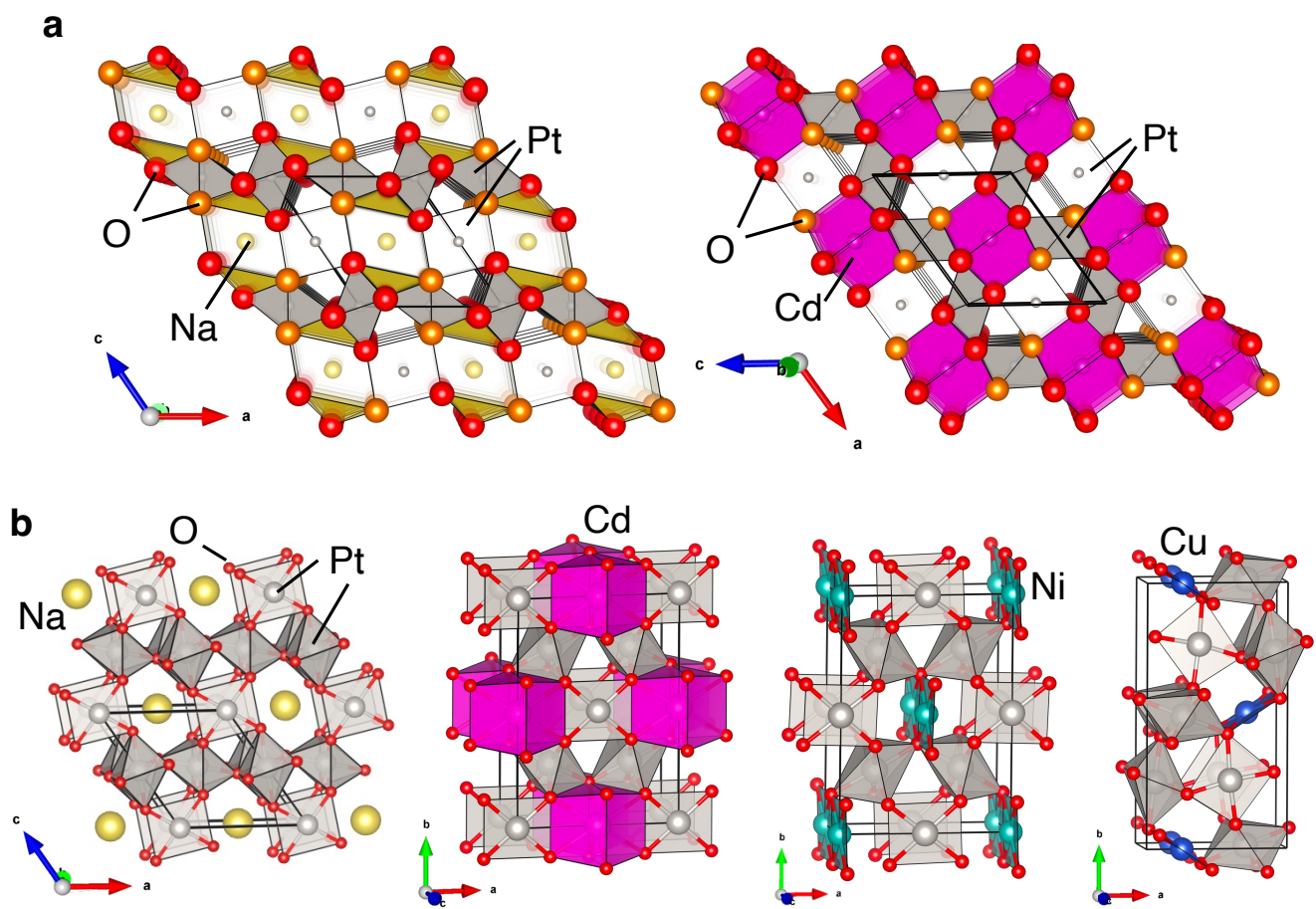

**Fig. S3. Structural polymorphism of $A$Pt$_3$O$_6$. a**, Crystal structures of NaPt$_3$O$_6$ and CdPt$_3$O$_6$, respectively. Each atom is shown as an ionic rigid sphere. For clarity, the primitive unit cell of CdPt$_3$O$_6$ is displayed instead of the conventional *Cmmm* unit cell shown below. The difference in the positions of oxygen is highlighted by orange, where the oxygen atoms are primarily modified by 0.5 unit along the *b*-axis. **b**, Crystal structures of NaPt$_3$O$_6$, CdPt$_3$O$_6$, NiPt$_3$O$_6$, and CuPt$_3$O$_6$, respectively. These structures were employed for the DFT calculations as initial parameters (see Fig. 2).

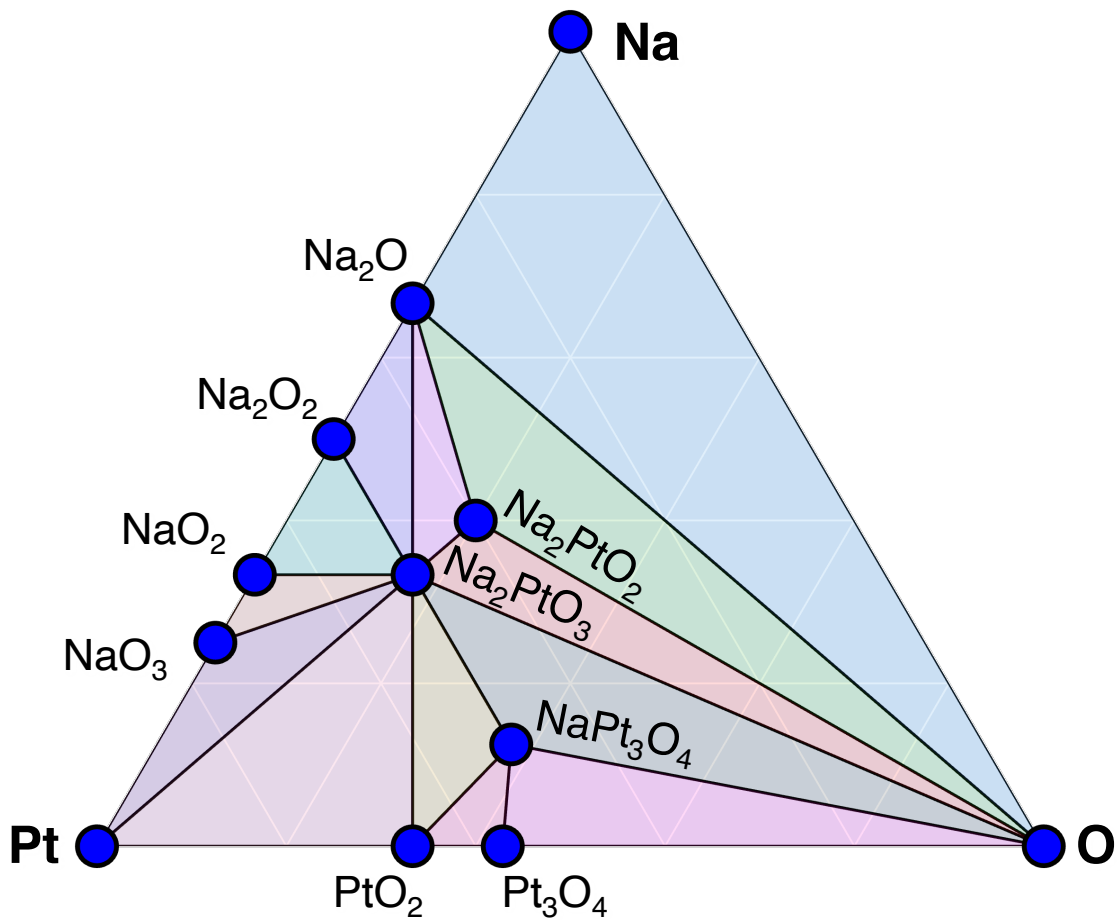

**Fig. S4. Convex hull diagram for Na-Pt-O ternary system under 0 GP.** Black solid lines show the convex hull. Blue circles show stable phases.

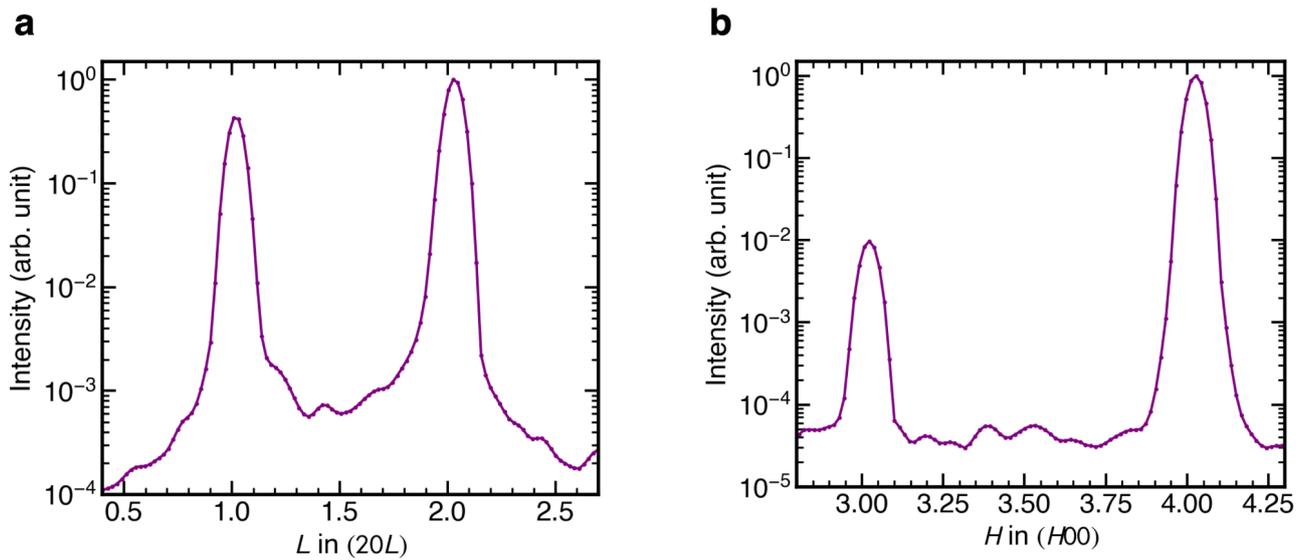

**Fig. S5. Intensity distributions in NaPt$_5$O$_{10}$ single-crystal SXRD data for the (a) 20$L$ and (b) $H$00 reflections.** Large background intensity corresponds to the diffuse scattering along the $c^*$ axis.

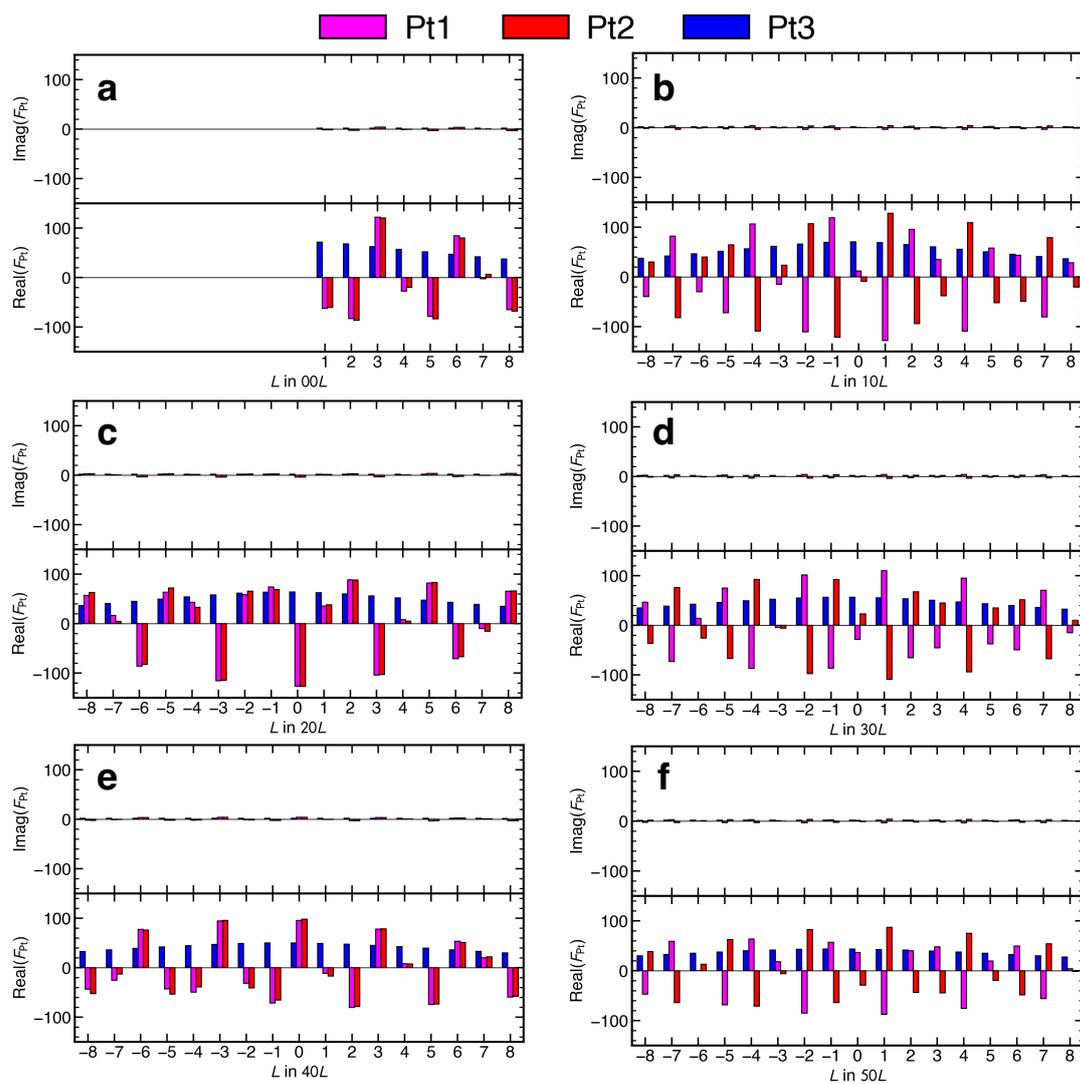

Fig. S6. Simulated real and imaginary parts of structural factors originated from Pt sites ($F_{Pt}$) for NaPt$_5$O$_{10}$. Structural factors were calculated along (a) 00$L$, (b) 10$L$, (c) 20$L$, (d) 30$L$, (e) 40$L$, and (f) 50$L$ reflections.

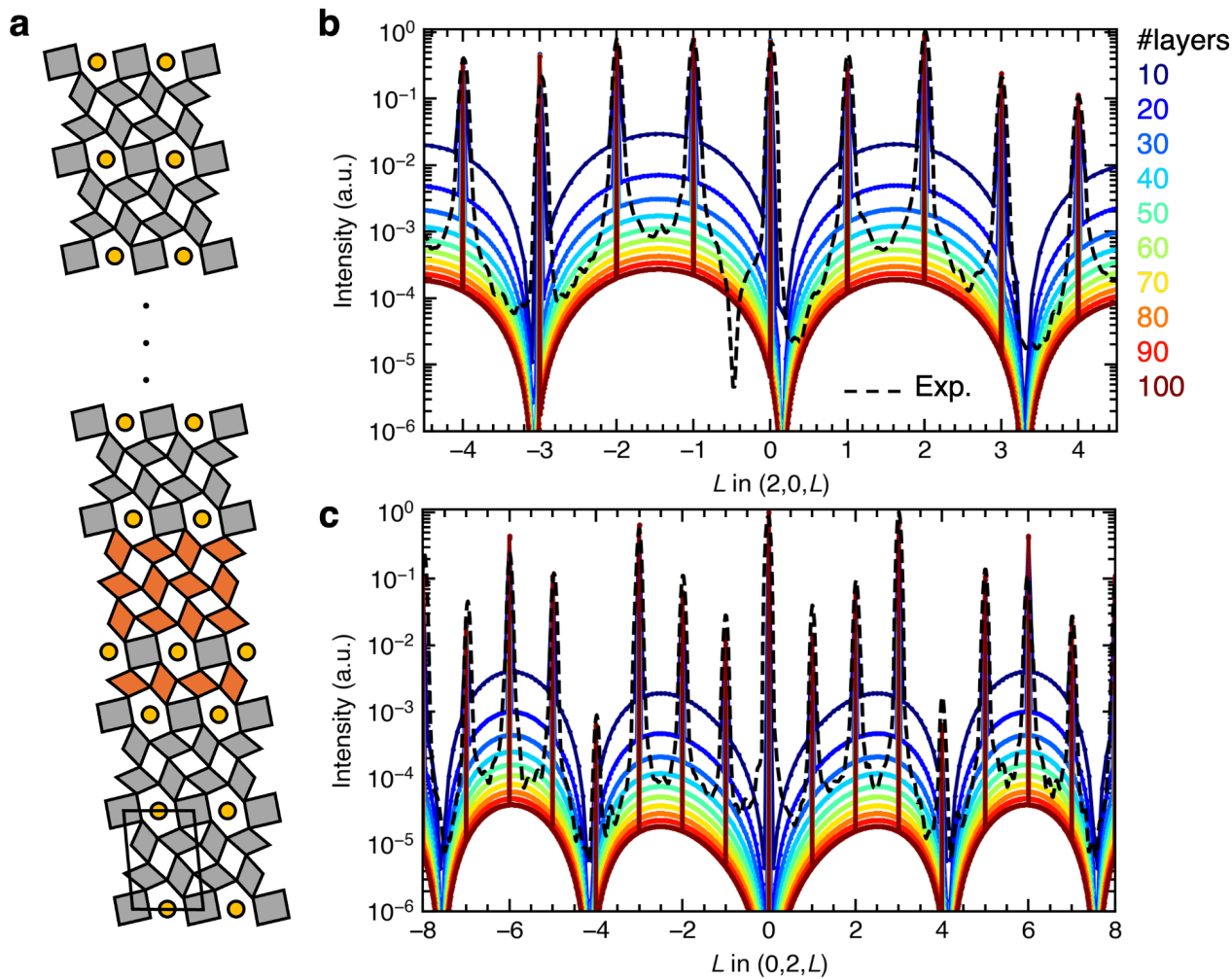

**Fig. S7. Diffuse scattering intensity distributions depending on the supercell size along the *c*-axis. a**, Schematic crystal structure of NaPt$_5$O$_{10}$ with varying number of PtO$_6$ layers used for simulation including 1-3-type stacking faults. Simulated and observed diffuse scattering intensity along (**b**) 20*L* and (**c**) 02*L* reflections.

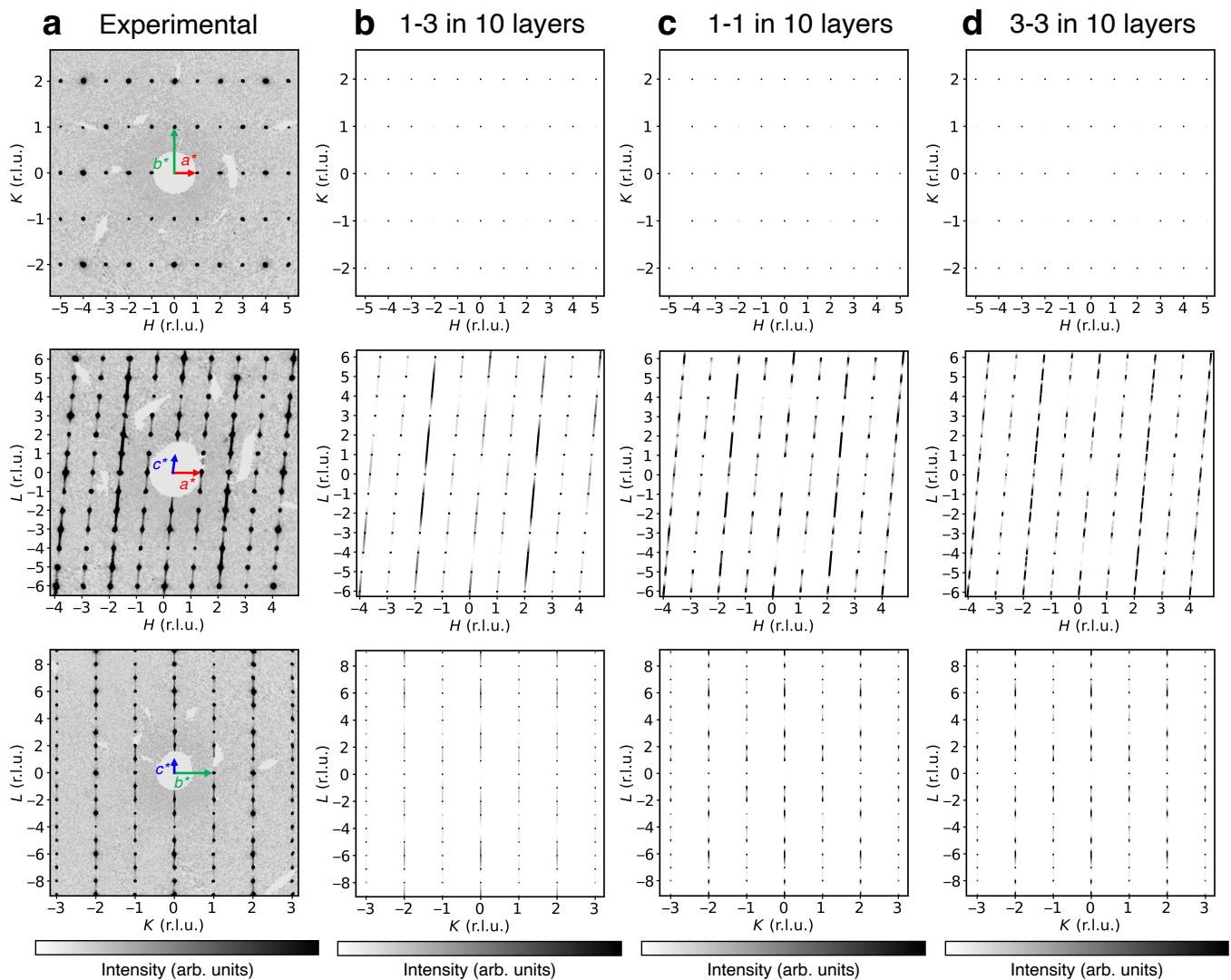

**Fig. S8. Effect of stacking fault patterns on simulated diffuse scattering extinction patterns. a**, Reconstructed precession images of the reciprocal lattice space obtained by SXRD experiments for the NaPt$_5$O$_{10}$ single crystal for $HK0$, $H0L$, and $0KL$ planes, respectively. Diffraction patterns simulated with 10-layer superstructures introducing (**b**) 1-3-type, (**c**) 1-1-type, and (**d**) 3-3-type stacking faults for $HK0$, $H0L$, and $0KL$ planes, respectively.

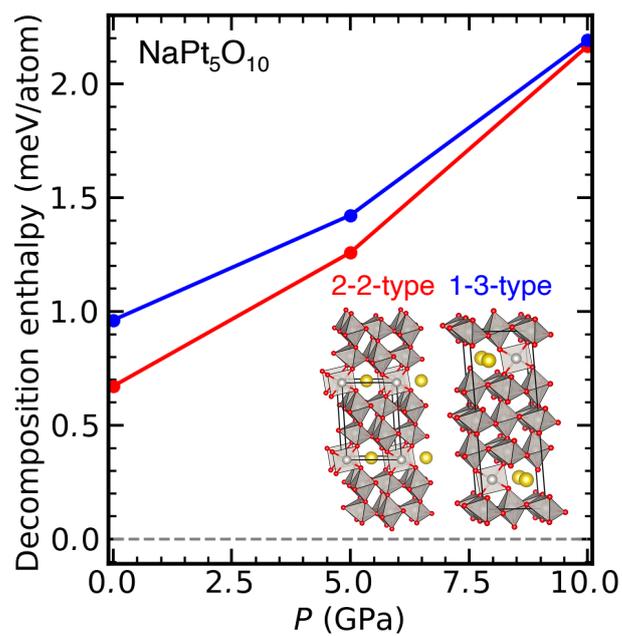

**Fig. S9. Pressure dependencies of decomposition enthalpies for 2-2-type and 1-3-type averaged structures of NaPt$_5$O$_{10}$ within the Na-Pt-O ternary convex hull.**

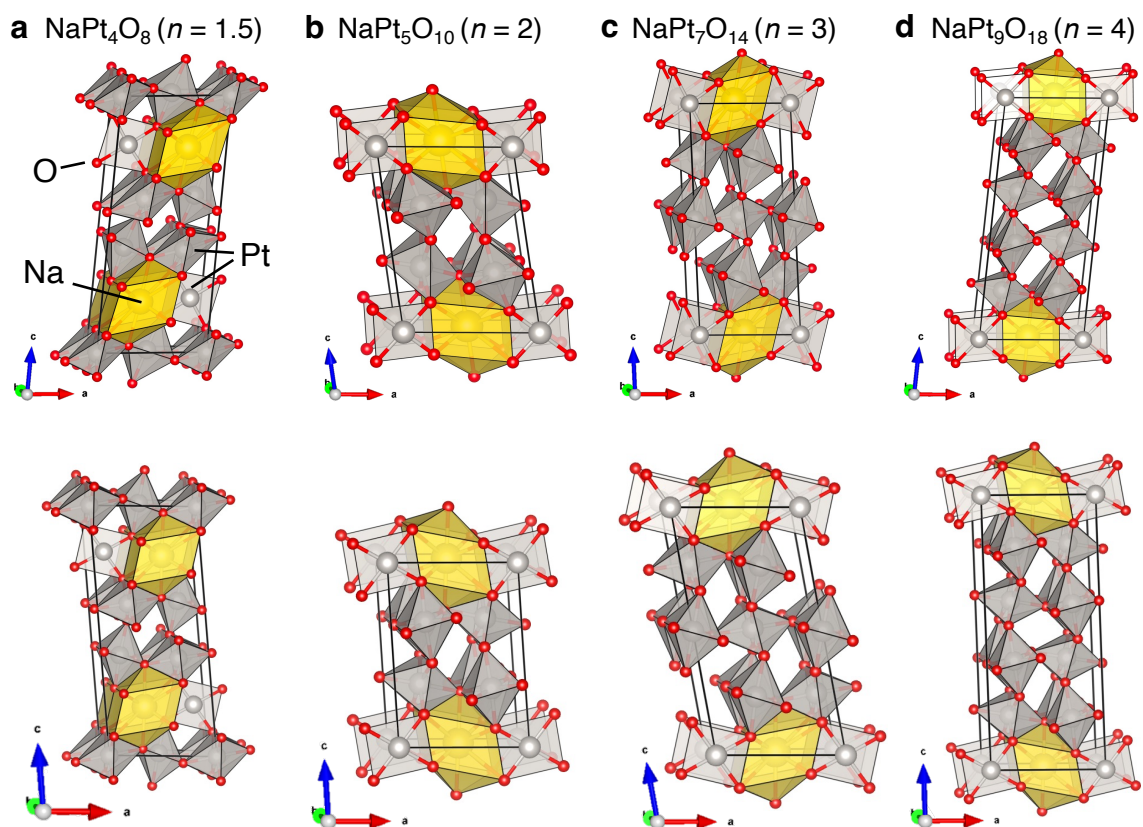

**Fig. S10. DFT-based structural relaxation of hypothetical Na(PtO$_2$)$_{2n+1}$ crystal structures.** (top) Manually generated crystal structures as initial configurations for (a) NaPt$_4$O$_8$, (b) NaPt$_5$O$_{10}$, (c) NaPt$_7$O$_{14}$, and (d) NaPt$_9$O$_{18}$. (bottom) Each optimized structure calculated by DFT.

**Table S1. Lattice parameters and atomic coordinates obtained from powder SXRD analysis of NaPt$_3$O$_6$.** The SXRD measurement was performed at 300 K using synchrotron radiation with a wavelength of λ = 0.6911 Å. The refinement yielded reliability factors of $R_{wp}$ = 1.37 and $S$ = 1.17. Occupancies of all sites are fixed at 1.0, and isotropic displacement parameters of oxygen sites are fixed at $B$ = 0.9 ($U_{iso}$ = 0.0114).

| Cell | a (Å) | b (Å) | c (Å) | β (°) | V (Å³) | |
|---|---|---|---|---|---|---|
| P2/m | 6.35(5) | 3.126(14) | 7.01(6) | 124.1(4) | 115(2) | |
| **Atom** | x | y | z | g | $U_{iso}$ (Å²) | Site |
| Na1 | 0.5 | 0.5 | 0 | 1 | 0.009(4) | 1f |
| Pt1 | 0.5 | 0.5 | 0.5 | 1 | 0.0023(4) | 1e |
| Pt2 | 0 | 0 | 0.5 | 1 | 0.0022(4) | 1g |
| Pt3 | 0 | 0 | 0 | 1 | 0.0027(5) | 1a |
| O1 | 0.774(2) | 0.5 | 0.433(2) | 1 | 0.0114 | 2n |
| O2 | 0.164(2) | 0 | 0.834(2) | 1 | 0.0114 | 2m |
| O3 | 0.637(2) | 0 | 0.704(2) | 1 | 0.0114 | 2m |

**Table S2.** Indexing and refinement statistics of the SXRD data of the NaPt$_3$O$_6$ single crystal.

| Data reduction | |
|---|---|
| – 0.28 Å resolution | |
| Completeness (%) | 99.3 |
| $R_{int}$ | 0.073 |
| Redundancy | 8.9 |
| $F^2/\sigma(F^2)$ | 25.9 |
| **Refinement** | |
| – Total 3576 unique reflections | |
| – 3281 are $F_o > 3\sigma(F_o)$ | |
| $R_1$ for $F_o > 3\sigma(F_o)$ | 3.43 |
| $R_1$ for all reflections | 4.06 |
| $wR_2$ for $F_o > 3\sigma(F_o)$ | 3.88 |
| $wR_2$ for all reflections | 4.01 |
| GoF for $F_o > 3\sigma(F_o)$ | 1.07 |
| GoF for all reflections | 1.05 |

**Table S3. Anisotropic displacement parameters (ADPs) refined using SXRD data of NaPt$_3$O$_6$ single crystal.** Each ADP was visualized in Fig. S1.

| ADP (Å$^2$) | $U_{11}$ | $U_{22}$ | $U_{33}$ | $U_{12}$ | $U_{13}$ | $U_{23}$ |
|---|---|---|---|---|---|---|
| Na1 | 0.0045(6) | 0.0019(7) | 0.0072(7) | 0 | 0.0013(6) | 0 |
| Pt1 | 0.00175(3) | 0.00233(5) | 0.00212(3) | 0 | 0.00119(2) | 0 |
| Pt2 | 0.00200(3) | 0.00255(5) | 0.00277(3) | 0 | 0.00150(2) | 0 |
| Pt3 | 0.00214(3) | 0.00326(5) | 0.00195(3) | 0 | 0.00122(2) | 0 |
| O1 | 0.0043(5) | 0.0038(6) | 0.0054(4) | 0 | 0.0035(3) | 0 |
| O2 | 0.0042(4) | 0.0056(6) | 0.0038(4) | 0 | 0.0026(3) | 0 |
| O3 | 0.0036(4) | 0.0039(6) | 0.0036(4) | 0 | 0.0014(3) | 0 |

**Table S4. Indexing and refinement statistics of the SXRD data of the NaPt$_5$O$_{10}$ single crystal.**

| Data reduction | |
|---|---|
| – 0.28 Å resolution | |
| Completeness (%) | 99.5 |
| $R_{int}$ | 0.025 |
| Redundancy | 5.3 |
| $F^2/\sigma(F^2)$ | 196.6 |
| **Refinement** | |
| – Total 8769 unique reflections | |
| – 8114 are $F_o > 3\sigma(F_o)$ | |
| $R_1$ for $F_o > 3\sigma(F_o)$ | 2.18 |
| $R_1$ for all reflections | 2.40 |
| $wR_2$ for $F_o > 3\sigma(F_o)$ | 2.90 |
| $wR_2$ for all reflections | 2.96 |
| GoF for $F_o > 3\sigma(F_o)$ | 1.59 |
| GoF for all reflections | 1.56 |

**Table S5. ADPs refined using SXRD data of NaPt$_5$O$_{10}$ single crystal.** Each atomic displacement was visualized in Fig. S1.

| ADP (Å$^2$) | $U_{11}$ | $U_{22}$ | $U_{33}$ | $U_{12}$ | $U_{13}$ | $U_{23}$ |
|---|---|---|---|---|---|---|
| Na1 | 0.0142(6) | 0.0095(5) | 0.0176(7) | 0 | –0.0067(5) | 0 |
| Pt1 | 0.00257(10) | 0.00245(10) | 0.002939(11) | 0 | 0.000116(7) | 0 |
| Pt2 | 0.00252(10) | 0.00261(10) | 0.003163(11) | 0 | 0.000414(8) | 0 |
| Pt3 | 0.003655(15) | 0.004581(16) | 0.002657(15) | 0 | 0.000221(11) | 0 |
| O1 | 0.00338(19) | 0.0041(2) | 0.0078(3) | 0 | 0.00076(19) | 0 |
| O2 | 0.0063(2) | 0.0044(2) | 0.0044(2) | 0 | –0.00102(19) | 0 |
| O3 | 0.0036(2) | 0.0039(2) | 0.0097(3) | 0 | 0.0008(2) | 0 |
| O4 | 0.0096(3) | 0.0039(2) | 0.0038(2) | 0 | 0.0006(2) | 0 |
| O5 | 0.0055(2) | 0.0101(3) | 0.0044(2) | 0 | 0.0013(2) | 0 |